\documentclass[sigconf]{acmart}
\renewcommand\footnotetextcopyrightpermission[1]{}
\usepackage{booktabs,multirow,amsmath,enumitem,graphicx,xcolor,hyperref,microtype,mathtools,subcaption,xspace}

\newcommand{\model}{\textsc{SIREN}\xspace}
\newcommand{\SemIDGSU}{\textsc{SIREN}\textsubscript{SemID-GSU}}
\newcommand{\SimGSU}{\textsc{SIREN}\textsubscript{Sim-GSU}}

\begin{document}

\title{\model: Unified Multi-Granularity Semantic Interaction for Multi-Modal Lifelong User Interest Modeling}

% ================= Author Info =================

% \author{
% Yaqian Zhang$^{1,*}$, 
% Ruyi Yu$^{1,*}$,
% Tianyi Li$^{2,*}$,
% Bohan Liu$^1$,
% Maoquan Ye$^1$,
% Ke Wang$^1$,
% Shifeng Wen$^{1,\dagger}$,
% Junwei Pan$^1$,
% Lijie Wang$^1$,
% Qi Zhou$^1$,
% Yeshou Cai$^1$,
% Chengguo Yin$^1$,
% Lifeng Wang$^1$,
% Hui Li$^2$,
% Lei Xiao$^1$,
% Haijie Gu$^1$
% }

\affiliation{
  \institution{
    Yaqian Zhang$^{1,*}$, 
    Ruyi Yu$^{1,*}$,
    Tianyi Li$^{2,*}$,
    Bohan Liu$^1$,
    Maoquan Ye$^1$,
    Ke Wang$^1$,
    Shifeng Wen$^{1,\dagger}$,
    Junwei Pan$^1$,
    Lijie Wang$^1$,
    Qi Zhou$^1$,
    Yeshou Cai$^1$,
    Chengguo Yin$^1$,
    Lifeng Wang$^1$,
    Hui Li$^2$,
    Lei Xiao$^1$,
    Haijie Gu$^1$
  }
  \institution{$^1$Tencent Inc., China \quad\quad $^2$School of Informatics, Xiamen University, China}
  \institution{\small$^1$\{yaqianzhang, sparkletyu, leobhliu, adamye, kirkkwang, romeowen, jonaspan, lijiewang, \\ joeyqzhou, showcai, turingyin, fandywang, shawnxiao, jerrickgu\}@tencent.com}
  \institution{$^2$litianyi@stu.xmu.edu.cn, hui@xmu.edu.cn}
  \country{$^*$Equal contribution. \quad $\dagger$Corresponding author.}
}

\renewcommand{\shortauthors}{Zhang et al.}

% ===============================================

\begin{abstract}
Industrial recommender systems increasingly leverage lifelong user behavior histories and rich multi-modal content to capture evolving user preferences. 
However, effectively integrating multi-modal features into lifelong interest modeling remains challenging due to the inherent misalignment between multi-modal and collaborative spaces. 
Existing paradigms typically rely on \emph{separate modeling} of multi-modal sequence and behavior sequence, and late fusion to alleviate the modality gap, which results in coarse-grained multi-modal representation and limited integration. 
In this paper, we propose \model, a \emph{unified multi-granularity semantic interaction framework} for multi-modal lifelong user interest modeling.
In the General Search Unit stage, we introduce two alternative retrieval strategies: multi-modal similarity-based soft retrieval for retrieval effectiveness, and Semantic ID (SemID)-based hard retrieval for efficient industrial serving. 
For the Exact Search Unit stage, we explicitly incorporate target-aware relevance via coarse similarity buckets and fine-grained prefix-encoded SemIDs, \emph{enabling unified interaction} with collaborative ID features within the target-conditioned transformer architecture.
Extensive experiments on the offline dataset demonstrate that \model achieves a state-of-the-art GAUC. 
Online A/B tests further demonstrate consistent GMV gains across multiple production scenarios, including +2.28\% in Weixin Moments, +3.87\% in Weixin Official Accounts, and +1.61\% in Weixin Channels.
From July 2025, \model has been fully launched for full-traffic serving in Tencent's advertising platform. 

\end{abstract}

\maketitle 

\section{Introduction}
% 介绍研究背景——长序列用户行为对推荐系统的重要性及时延挑战
Lifelong user behavior histories provide rich personalized signals and play a crucial role in industrial recommender systems, especially in advertising, feed, and e-commerce scenarios~\cite{UBM,tencent-long}.
Modeling user interests over such long-term histories has been shown to substantially improve click-through rate (CTR) prediction by capturing persistent, diverse, and evolving user preferences~\cite{zhou2018din,pi2020sim,ETA,twin,twinv2,dare}. Nevertheless, directly leveraging lifelong behaviors is computationally challenging, due to their extreme length and the strict latency constraints in online systems~\cite{pi2020sim,ETA,tencent-long}.

% 介绍现有长序列建模的两种主流范式，并强调两阶段方法的工业优势
Existing long-sequence recommendation methods can be broadly categorized into two paradigms.
The first line of work attempts to model long behavioral sequences end-to-end by designing efficient attention or compression mechanisms~\cite{longer,stca,ultra-hstu}.
The second line adopts a two-stage paradigm~\cite{pi2020sim,twin,ETA,dare}, where the first General Search Unit (GSU) stage retrieves a short target-relevant subsequence from the full behavior history, and the second Exact Search Unit (ESU) stage performs user interest modeling over the retrieved behaviors.
Due to its favorable efficiency and deployment flexibility, the two-stage paradigm has become a widely adopted solution in large-scale industrial recommendation systems.

\begin{figure}[!t]
    \centering
    %\vspace{10pt}
    % 左侧子图 (a)
    \begin{subfigure}[t]{0.23\textwidth}
        \centering
        \includegraphics[width=\textwidth]{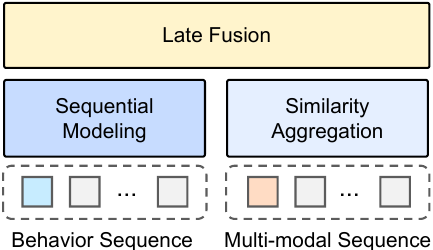}
        \caption{Separate Modeling Paradigm}
        \label{fig:teaser_a}
    \end{subfigure}
    \hfill
    % 右侧子图 (b)
    \begin{subfigure}[t]{0.23\textwidth}
        \centering
        \includegraphics[width=\textwidth]{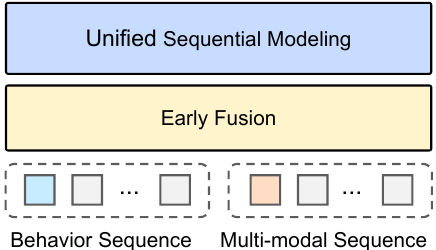}
        \caption{Unified Modeling Paradigm}
        \label{fig:teaser_b}
    \end{subfigure}
    \caption{
Comparison between separate and unified paradigms for multi-modal lifelong user interest modeling.
Existing methods usually encode ID-based behavior sequences and multi-modal sequences separately, followed by late fusion after sequence aggregation, which limits fine-grained interaction between collaborative and semantic signals.
SIREN instead performs item-level early fusion, allowing multi-modal semantics and collaborative features to interact within a unified sequential modeling framework.
}
    % \caption{Comparison between two modeling paradigms for multi-modal user interest modeling.}
    \label{fig:teaser}
    \vspace{-15pt}
\end{figure}

% 引出多模态信息在推荐中的潜力及其与协同空间不匹配的问题
Recent advances in multi-modal foundation models enable high-quality content representation from diverse item modalities~\cite{Clip,MMLLMSurvey,Qwen3-vl}, motivating their adoption in lifelong user interest modeling~\cite{SimTier,MISS,muse}. Compared with ID-based signals, multi-modal semantics provide stronger generalization ability in long-tail and cold-start scenarios. 
However, effectively integrating multi-modal signals into recommender systems remains challenging due to the misalignment between multi-modal and collaborative spaces~\cite{MISSRec,AlignRec}. 
Specifically, pre-trained multi-modal embeddings mainly capture content similarity rather than collaborative signals, and their distributions are often incompatible with ID embeddings, making naive integration prone to introducing noise and degrading recommendation performance~\cite{MMRecSurvey1,MMRecSurvey2,MISSRec}.

\begin{figure}[t]
\centering
\includegraphics[width=0.50\textwidth]{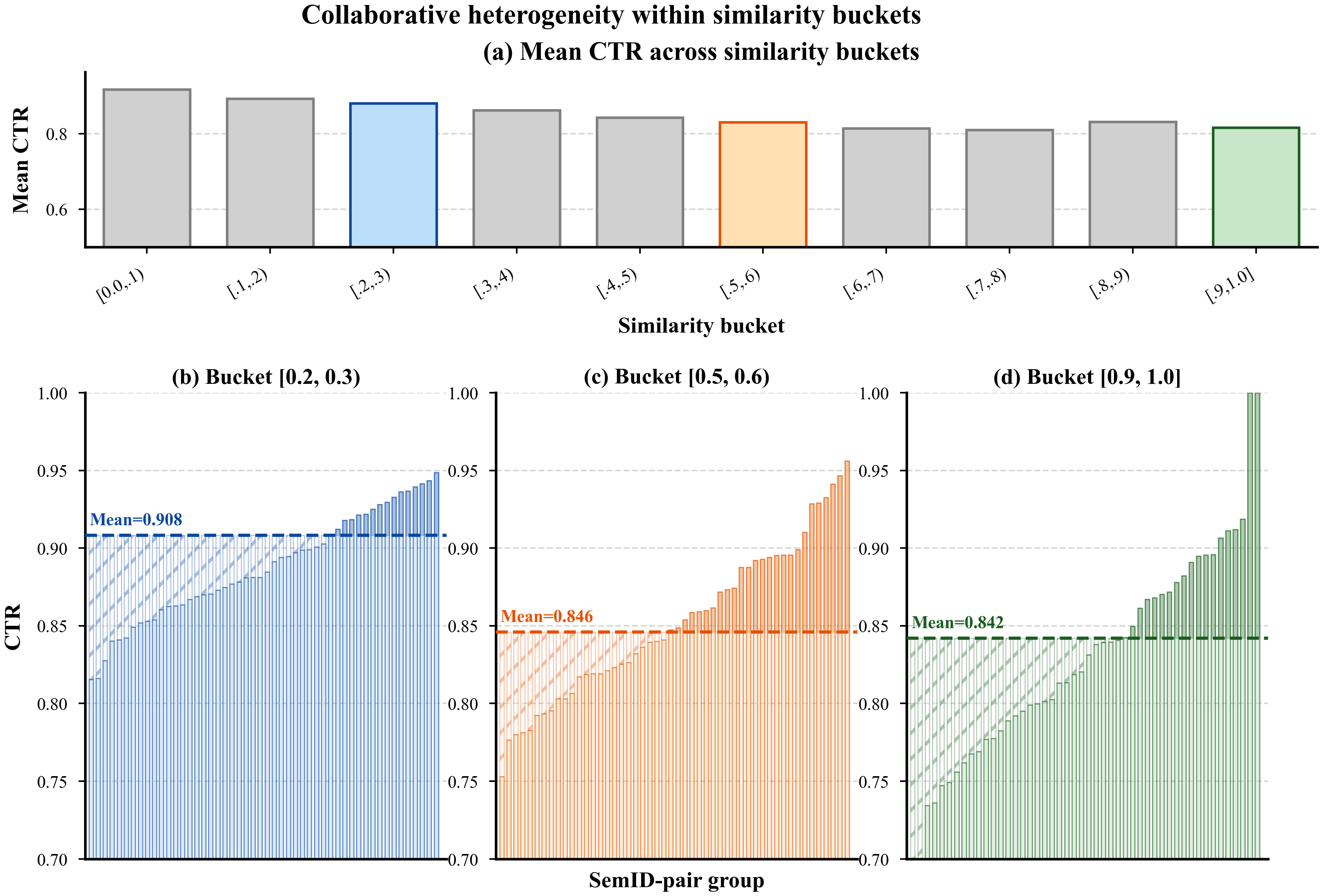}
\caption{
CTR distributions of Semantic ID groups within target--behavior similarity buckets.
Although pairs in the same bucket have similar multi-modal proximity to the target item, their CTRs vary substantially across Semantic ID groups, especially in high-similarity regions.
This within-bucket heterogeneity shows that similarity buckets provide only coarse target-aware relevance, motivating the use of fine-grained SemIDs to preserve semantic and collaborative distinctions.
}
% \caption{CTR distributions of different SemID groups within target--behavior similarity buckets.}
\label{fig:teaser_coarsefine}
\vspace{-10pt}
\end{figure}

To bridge this gap, existing studies typically adopt a separate modeling paradigm, where multi-modal sequences and behavior sequences are modeled individually, and a late fusion process is applied, as illustrated in Fig.~\ref{fig:teaser}(a). Despite the effectiveness, the prevailing separate modeling paradigm suffers from two issues:
\begin{enumerate}[leftmargin=*]
\item \textbf{Multi-modal signals and collaborative signals are not aligned within a unified representation space, which restricts feature interaction.} 
Prior methods~\cite{muse} typically treat multi-modal information as an auxiliary branch and fuse it with ID-based user representations only after sequence aggregation. 
\emph{Such late-stage fusion prevents multi-modal signals from sufficiently interacting with collaborative signals during sequential modeling}. 
As a result, multi-modal information mainly serves as attention modulation or sequence-level augmentation~\cite{SimTier,muse}, rather than directly participating in item-level representation learning, thereby limiting the model's ability to capture more discriminative user interest representations.

\item \textbf{Target--behavior similarity provides only a coarse-grained view and overlooks collaborative heterogeneity within multimodal proximity.}
Although target--behavior similarity is effective for target-aware multi-modal interest extraction, it compresses the rich behavior--target relationship into a scalar or bucketized proximity measure~\cite{SimTier}. 
This coarse representation implicitly treats behavior--target pairs with similar multi-modal proximity as similarly informative for predicting user responses. 
However, Fig.~\ref{fig:teaser_coarsefine} shows that pairs within the same similarity bucket can still exhibit markedly different CTR patterns across Semantic ID groups, especially in high-similarity regions. 
This indicates that \emph{multi-modal proximity captures coarse target relevance, but does not preserve the collaborative structure that governs user feedback}. 
Therefore, similarity-based representations alone struggle to distinguish behavior--target pairs that are close in the multi-modal space yet behave differently in the collaborative CTR space, limiting their ability to model fine-grained user interests.
\end{enumerate}

% \item \textbf{Target--behavior similarity provides only a coarse-grained representation and is insufficient to model fine-grained item heterogeneity.} 
% While target--behavior similarity is effective for target-aware multi-modal interest extraction, it inevitably compresses rich item semantics into scalar or aggregated similarity representations~\cite{SimTier}. 
% As illustrated in Fig.~\ref{fig:teaser_coarsefine}, even behaviors within the same similarity bucket can exhibit substantially different CTR patterns across Semantic ID groups, and this heterogeneity becomes more pronounced in high-similarity regions. 
% This observation suggests that \emph{similarity primarily captures coarse target relevance, while overlooking fine-grained semantic distinctions between behaviors}. 
% Consequently, \emph{similarity-based representations alone struggle to distinguish behaviors with comparable relevance scores but different semantic content}, limiting the model's ability to model fine-grained user interests.
% \end{enumerate}

% 介绍本文方法及其核心思想（统一融合语义ID与相似度信号）
To address these limitations, we propose \model, a unified multi-granularity framework that bridges multi-modal semantics and collaborative user interests for lifelong user modeling. 
Following the industrial GSU-ESU paradigm, \model incorporates multi-modal signals into both retrieval and exact interest modeling. 
In GSU, \model provides two complementary retrieval mechanisms: similarity-based soft retrieval, which uses dense multi-modal embeddings to retrieve target-relevant behaviors, and SemID-based hard retrieval, which uses discrete Semantic IDs as retrieval keys to enable efficient lookup in large-scale serving systems. 
In ESU, \model introduces a unified \emph{target-conditioned Transformer}~\cite{tin,HSTU,dare} for explicit behavior--target interaction. 
Instead of fusing separately modeled multi-modal and collaborative representations at a late stage, \model performs item-level early fusion by combining ID features, coarse target-aware similarity buckets, and fine-grained prefix-encoded SemIDs within the same target-conditioned backbone. 
Importantly, these signals are not only used to modulate attention weights, but are also integrated into behavior representations before sequence aggregation, \emph{allowing behavior--target interactions to shape both attention and representation learning}. 
As a result, \model jointly captures collaborative dependencies, coarse multi-modal relevance, and fine-grained semantic heterogeneity in a unified interaction representation space.

We conduct extensive experiments on the Taobao-MM dataset and online A/B tests in the Weixin advertising system.
Offline results show that \model consistently outperforms strong baselines in terms of GAUC, achieving the best GAUC with a relative improvement of +2.48\%.
Further analyses demonstrate that \model learns more discriminative user representations and better captures semantic heterogeneity beyond similarity-based methods.
For online evaluation, \model achieves consistent GMV improvements across multiple Weixin production scenarios, with gains of +2.28\% in Moments, +3.87\% in Official Accounts, and +1.61\% in Channels.
The improvements are more pronounced in cold-start settings, including low-activity users and newly launched ads.
These results validate the effectiveness and robustness of \model in real-world large-scale deployment.

In summary, our contributions are as follows: 
\begin{itemize}[leftmargin=*]
    \item We propose a lifelong user interest modeling framework that effectively incorporates multi-modal signals into the industrial GSU--ESU two-stage paradigm.

    \item We investigate both multi-modal similarity-based soft retrieval and SemID-based hard retrieval in the GSU stage, providing a practical trade-off between retrieval quality and online serving efficiency.
    
    \item We introduce target-aware similarity buckets and prefix-encoded SemID as complementary multi-modal features for ESU modeling, enabling fine-grained early fusion with ID-based collaborative features upon a target-conditioned Transformer.
    
    \item Extensive offline experiments, representation analyses and online A/B tests demonstrate the superior effectiveness of \model in large-scale industrial scenarios.
\end{itemize}
\section{Related Work}
\subsection{Lifelong User Interest Modeling}
Modeling long-term user interests from lifelong behavior sequences is a fundamental challenge in industrial recommender systems. To address the scalability issue of ultra-long sequences, industrial recommenders widely adopt a two-stage paradigm (GSU and ESU) introduced by SIM~\cite{pi2020sim}. Subsequent works further improve retrieval efficiency and scalability through techniques such as locality-sensitive hashing~\cite{ETA}, decoupled representation learning~\cite{twin}, hierarchical sequence compression~\cite{twinv2}, and decoupled embeddings for retrieval and ranking objectives~\cite{dare}.
More recently, several studies have explored end-to-end long-sequence modeling by redesigning attention mechanisms and improving computational efficiency~\cite{longer,stca,tencent-long,lcn}. Despite their strong performance, these methods are still predominantly based on ID-centric representations, which often suffer from semantic cold-start issues and limited cross-domain generalization ability.

\subsection{Multi-Modal Sequential Recommendation}
Multi-modal signals provide semantic information beyond ID-based collaborative signals, but their integration into sequential recommendation remains challenging due to the misalignment with collaborative spaces. One line of research focuses on semantic tokenization and alignment via semantic ID learning or quantization~\cite{QARM,letter,DAS,MMQ}, while paying limited attention to how semantic signals are explicitly incorporated into sequential interest modeling.
Another line of work utilizes target-aware multi-modal similarity to enhance sequence modeling. SimTier~\cite{SimTier} transforms target--behavior multi-modal similarities into histogram-style representations. The state-of-the-art lifelong interest modeling method MUSE~\cite{muse} further introduces Semantic-Aware Target Attention (SA-TA), which enhances target-aware interest extraction by integrating multi-modal similarity signals with behavior attention scores. Nevertheless, existing methods typically rely on coarse sequence-level aggregation and late fusion, limiting fine-grained interactions between semantic and collaborative features.

\section{Preliminaries}
\label{sec:preliminary}

% \subsection{Problem Formulation}
We study lifelong user interest modeling for CTR prediction in a multi-modal setting. 
Given contextual information $c$, a user $u$, and a target item $v_t$, let 
$H=(b_1,\dots,b_N)$ denote the full behavior history of length $N$. 
Following the industrial two-stage paradigm, the General Search Unit (GSU) first retrieves a target-relevant subsequence 
$H_t=(b_1,\dots,b_L)$ with $L\ll N$, and the Exact Search Unit (ESU) then performs fine-grained interest modeling over $H_t$.

Each item $b_i$ is represented by $x_i=(\mathbf{z}_i^{id}, e_i^{mm})$, where $\mathbf{z}_i^{id}$ denotes ID-based categorical features and $e_i^{mm}\in\mathbb{R}^{d}$ is a pre-trained multi-modal embedding aligned with collaborative signals through large-scale user interactions. 
The objective is to learn a predictive function
$\hat{y}_t=P(y_t=1\mid H_t,v_t,u,c)$ by minimizing the binary cross-entropy loss with the ground-truth label $y_t\in\{0,1\}$.

% =============================================================
% The SIREN Framework: GSU + ESU
% =============================================================
\section{Overall Architecture}
\label{sec:siren}

\begin{figure}[t]
\centering
\includegraphics[width=0.5\textwidth]{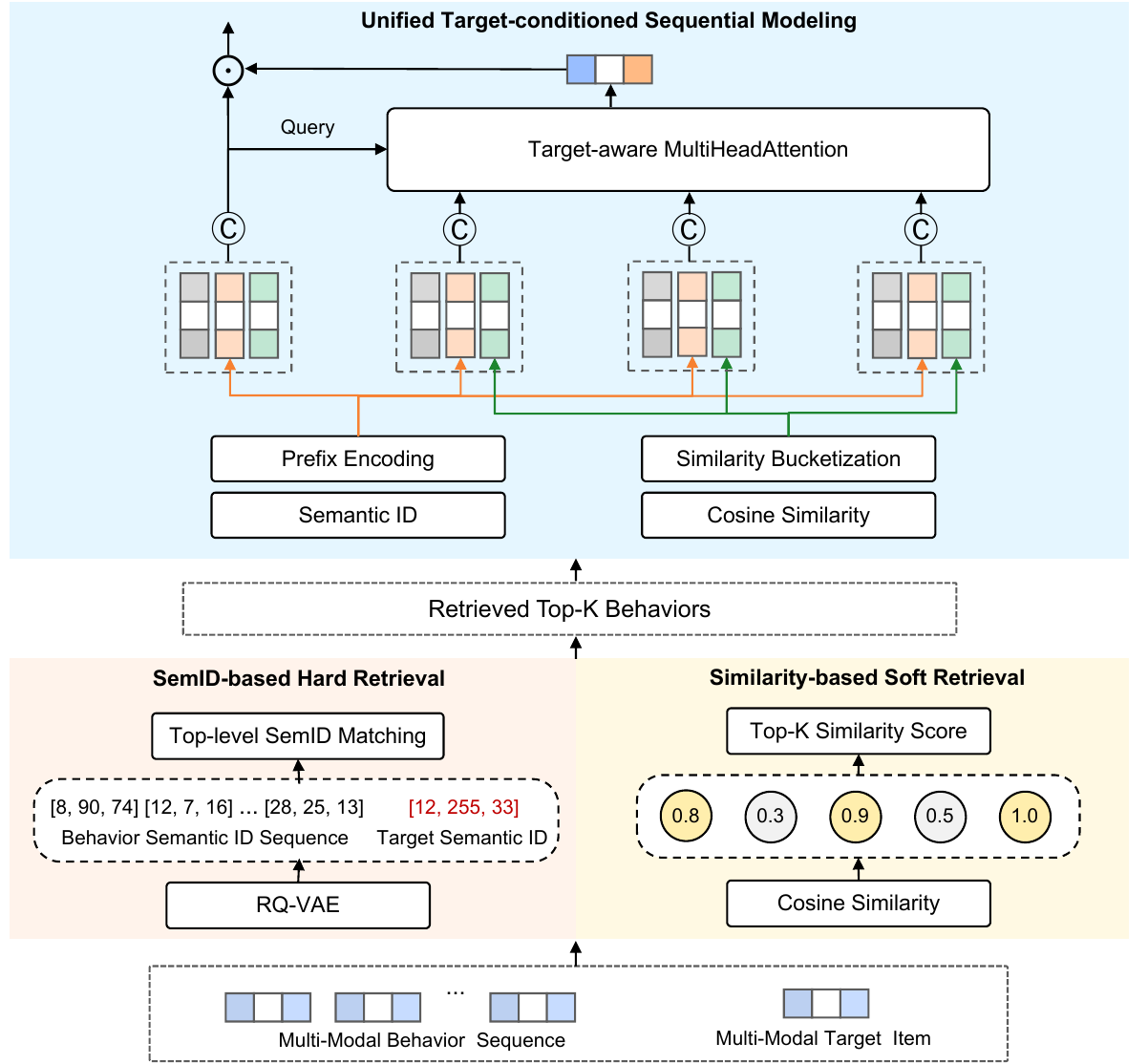}
\caption{Overview of \model.
\model\ follows the two-stage GSU–ESU paradigm. The GSU retrieves target-relevant behaviors via similarity-based soft retrieval or SemID-based hard retrieval. The ESU incorporates SemIDs and similarity buckets as side information for unified target-conditioned sequence modeling.}
\label{fig:architecture}
\end{figure}

In this section, we first introduce the multi-modal feature construction in SIREN, and then present the overall two-stage architecture:
% In this section, we detail the multi-modal feature construction, the \model\ architecture, which follows a two-stage paradigm:

\begin{itemize}[leftmargin=*]
    \item \textbf{GSU:} We introduce two multi-modal retrieval strategies, namely \textit{similarity-based soft search} and \textit{SemID-based hard search}—to provide a trade-off between retrieval quality and deployment efficiency.
    \item \textbf{ESU:} We present a unified sequential modeling framework that seamlessly incorporates SemIDs and similarity buckets into sequence modeling, enriched by target-conditioned interaction.
\end{itemize}

\subsection{Multi-modal Feature Construction}
\label{subsec:feature_construction}
To enable unified multi-modal interest modeling, \model\ constructs two complementary feature types from pre-trained multi-modal embeddings: SemID and target-aware similarity bucket.
\subsubsection{Semantic ID Construction}
\label{sec:SemID}

To make continuous multi-modal embeddings compatible with ID-centric recommendation systems, we transform each item embedding into a discrete Semantic ID (SemID) using RQ-VAE~\cite{RQ-VAE}. 
Given $e_i^{mm}\in\mathbb{R}^{d}$, RQ-VAE produces a hierarchical code sequence:
\begin{equation}
    \mathrm{SemID}_i = \big(c_i^{(1)}, c_i^{(2)}, \dots, c_i^{(M)}\big),
\end{equation}
where different codes capture semantic information at different quantization levels, forming a coarse-to-fine representation of the item.

Rather than embedding each code independently, we adopt prefix encoding~\cite{prefix-sid}. 
For $\mathrm{SemID}_i$, we construct prefix tokens
\begin{equation}
    \mathcal{P}_i =
    \left\{
    c_i^{(1)},
    (c_i^{(1)},c_i^{(2)}),
    \dots,
    (c_i^{(1)},\dots,c_i^{(K)})
    \right\},
\end{equation}
where $K\le M$ is the maximum prefix depth. 
Each prefix $p\in\mathcal{P}_i$ is mapped to a learnable embedding via a shared lookup table, and the final semantic representation is
\begin{equation}
    e_i^{sem} =
    \mathrm{Concat}\big(\{\mathbf{E}_{\mathrm{prefix}}[p]\mid p\in\mathcal{P}_i\}\big).
\end{equation}
This prefix-based representation preserves hierarchical multi-modal semantics while providing a discrete interface for both efficient GSU retrieval and fine-grained ESU representation.

\subsubsection{Target-aware Similarity Bucketization}
\label{sec:bucket}

While SemIDs encode item semantics, ESU modeling also requires explicit target-conditioned relevance between each historical behavior and the target item. 
For each behavior item $b_i$ and target item $v_t$, we compute their multi-modal cosine similarity:
\begin{equation}
    s_{i,t}=\mathrm{sim}(e_i^{mm},e_t^{mm})
    =\frac{(e_i^{mm})^\top e_t^{mm}}
    {\|e_i^{mm}\|\cdot\|e_t^{mm}\|}.
\end{equation}
The continuous score is then discretized into a bucket index:
\begin{equation}
    q_{i,t}=\mathcal{B}(s_{i,t})
    =\left\lfloor
    \frac{s_{i,t}-s_{\min}}{s_{\max}-s_{\min}}\cdot B
    \right\rfloor,
\end{equation}
where $B$ is the number of buckets, and $[s_{\min},s_{\max}]$ is the effective similarity range estimated from data statistics. 
Values outside this range are clipped to boundary buckets. 
Each bucket index is mapped to a learnable embedding:
\begin{equation}
    e_{i,t}^{Sim}=\mathrm{Emb}^{sim}(q_{i,t}).
\end{equation}

\subsection{General Search Unit}
\label{sec:gsu}
A key requirement of the GSU stage is to balance retrieval efficiency with relevance quality when operating over ultra-long sequences. 
Conventional ID-based retrieval strategies suffer from limited semantic expressiveness and fail to capture content-level relevance.
To address these limitations, we explore two multi-modal retrieval strategies to enhance the quality of candidate behavior:

\paragraph{\textbf{Similarity-based Soft Retrieval.}} Following ~\cite{muse}, we retrieve the top-$K$ behaviors based on the cosine similarity between the multi-modal embeddings of historical items and the target item:
\begin{equation}
    \mathcal{S}_{\text{sim}} = \operatorname{Top\text{-}K}_{b_i \in \mathcal{B}} 
    \;\text{sim}(e_i^{mm},\, e_t^{mm}),
    \label{eq:gsu_soft}
\end{equation}
where $\text{sim}(\cdot,\cdot)$ denotes cosine similarity.
While effectively capturing dense content relevance, this strategy incurs heavy online computational overhead due to full-sequence similarity matching. Moreover, maintaining real-time multi-modal embedding indices introduces significant storage and system complexity, limiting its scalability in large-scale production systems.

\paragraph{\textbf{Semantic ID-based Retrieval.}}
Driven by these limitations, we further explore another retrieval strategy, where SemIDs serve as efficient retrieval keys. 
Specifically, we use the top-level semantic code $c_t^{(1)}$ of the target item to query an inverted index, retrieving all historical behaviors that share an identical top-level code:
\begin{equation}
    \mathcal{S}_{\text{SemID}} = \big\{\, b_i \in \mathcal{B} \;\big|\; 
    c_i^{(1)} = c_t^{(1)} \,\big\}.
    \label{eq:gsu_hard}
\end{equation}

Compared to soft retrieval, this strategy offers two key advantages. First, it replaces dense similarity computations with inverted index lookups, reducing online complexity from $O(|\mathcal{B}| \cdot d)$ to near-constant time. Second, it eliminates storing and transmitting high-dimensional embeddings, significantly lowering memory and bandwidth overhead.

\subsection{Exact Search Unit}
\label{sec:esu}
The ESU stage is responsible for fine-grained sequence modeling over the retrieved behaviors. In order to introduce multi-modal signals, a central challenge lies in the misalignment between multi-modal embeddings and ID-based signals. 

To address this, we leverage the multi-modal features introduced in Sec .~\ref{subsec:feature_construction} as side information. 
This information is integrated at the item level within the unified target-conditioned modeling framework, enabling fine-grained interaction between historical behaviors and the target item. Fig.~\ref{fig:architecture} illustrates the overall pipeline.

\paragraph{\textbf{Unified Item Representation}}
\label{sec:unified}

For each behavior item $b_i$ in the sequence $H_t$, we construct a unified representation by combining collaborative and multi-modal features:
\begin{equation}
    h_i = \text{Concat}\big(e_i^{id},\; e_i^{sem}\big),
\end{equation}
where $e_i^{id}$ denotes the concatenated embeddings of $\mathbf{z}_i^{id}$.

The target item $v_t$ is represented in the same form, i.e., $h_t = \text{Concat}(e_t^{id}, e_t^{sem})$.

\paragraph{\textbf{Target-aware Sequential Modeling.}}
Given the unified representations, we employ a target-aware sequence encoder to extract user interest:
\begin{equation}
    u_t = f(\{h_1, \dots, h_L\}, h_t),
\end{equation}
where $f(\cdot)$ denotes a sequence modeling function. Specifically, $f$ is instantiated as a multi-head target attention mechanism, where the importance of each behavior is conditioned on the target item:
\begin{equation}
\label{eq:user_interest}
    \alpha_i = \text{Attn}\big(h_i \oplus e_{i,t}^{Sim},\; h_t \oplus e_t^{Sim}\big), \qquad
    u_t = \sum_{i=1}^{L} \alpha_i h_i,
\end{equation}
where $e_{i,t}^{Sim}$ is the similarity bucket embedding, and $e_t^{Sim}$ is a learnable embedding associated with the target item. $\alpha_i$ denotes the attention weight for behavior $b_i$, $\oplus$ denotes vector concatenation.

A key property of this design is that all feature types---including ID-based, semantic, and target-aware signals---are integrated prior to sequence aggregation. 
Consequently, these signals jointly influence both attention weights and behavior representations during interest extraction.

\paragraph{\textbf{Target-conditioned Interaction.}}
To further align the extracted user interest with the target item, we introduce a lightweight element-wise interaction in the representation space~\cite{tin, HSTU, dare}:
\begin{equation}
\label{eq:target_aware_user_interest}
    \tilde{u}_t = u_t \odot h_t,
\end{equation}
where $\odot$ denotes element-wise multiplication.

The resulting representation is used for final prediction:
\begin{equation}
    \hat{y}_t = \sigma\big(g([\tilde{u}_t, h_t, u, c])\big).
\end{equation}

This interaction captures fine-grained correlations between user interest and target representations, allowing the model to capture fine-grained compatibility patterns beyond attention-based weighting.

\section{Experiments}
\label{sec:exp}

\subsection{Experimental Setup}
\label{sec:setup}

\subsubsection{Dataset.}
We evaluate our method on the Taobao-MM dataset\footnote{\url{https://huggingface.co/datasets/TaoBao-MM/Taobao-MM}}, a large-scale public benchmark for multi-modal lifelong user behavior modeling~\cite{muse}. 
The dataset is collected from real-world traffic of Taobao's display advertising system and contains long-term user behavior sequences paired with high-quality multi-modal representations.
Each item is associated with standard ID-based categorical features as well as a 128-dimensional pre-trained multi-modal embedding generated by SCL~\cite{muse}. 

The dataset contains 99M interaction samples from 8.79M users over 35.4M items. 
Each user is associated with a historical behavior sequence of up to 1K interactions, making it a realistic benchmark for evaluating long-sequence recommendation methods. 
Each sample consists of anonymized user features (e.g., user ID, age, gender, location), item features (e.g., item ID, category), and a binary click label.

\subsubsection{Baselines.}
We compare \model against a representative set of lifelong user interest modeling methods.
\begin{itemize}[leftmargin=*]
    \item \textbf{DIN}~\cite{zhou2018din}: A target attention model that operates on short-term user behavior sequences without a GSU retrieval stage.
    \item \textbf{SIM-Hard}~\cite{pi2020sim}: A two-stage model where the GSU retrieves behaviors based on exact category match with the target item.
    \item \textbf{SIM-Soft}~\cite{pi2020sim}: A variant of SIM where the GSU retrieves behaviors based on inner-product similarity of item embeddings.
    \item \textbf{TWIN}~\cite{twin}: A two-stage model that ensures consistency between GSU and ESU by adopting the same target attention mechanism in both stages.
    \item \textbf{MISS}~\cite{MISS}: A multi-modal enhanced retrieval method that introduces a multi-modal GSU alongside the ID-based one.
    \item \textbf{MUSE}~\cite{muse}: A state-of-the-art multi-modal lifelong interest modeling framework that integrates multi-modal signals into both GSU and ESU stages.
    \item \textbf{\SemIDGSU}: A variant of the \model where the GSU stage uses Semantic ID-based hard retrieval, while keeping the ESU unchanged.
    \item \textbf{\SimGSU}: A variant of \model where the GSU stage adopts similarity-based soft retrieval, while keeping the ESU unchanged.
\end{itemize}

\subsubsection{Implementation Details.}
For a fair comparison, each model is trained for one epoch following standard practice~\cite{muse, twin}. For DIN, we use the most recent 50 behaviors, as it is a single-stage model without a GSU retrieval. For all two-stage models (SIM, TWIN, MISS, MUSE, \model), the GSU retrieves top-50 behaviors from the lifelong behavior sequence for ESU modeling. For SIM-Hard and SIM-Soft, we adopt the original GSU design while replacing the ESU with the \model implementation. By default, \model employs similarity-based soft retrieval in the GSU stage. We use AdamW~\cite{AdamW} for dense parameters and SparseAdam for sparse embedding parameters, with learning rates set to $2\times10^{-4}$ and $2\times10^{-3}$, respectively. The batch size is set to 1000. We adopt Group AUC (GAUC) as the primary offline evaluation metric.

\subsection{Overall Performance}
\label{sec:overall}

Table~\ref{tab:overall} reports the offline performance of all methods. We make the following observations:

\begin{table}[t]
\centering
\setlength{\tabcolsep}{14pt}
\caption{Overall offline performance comparison on GAUC. The best result is highlighted in \textbf{bold}.}
\label{tab:overall}
\begin{tabular}{lcc}
\toprule
\textbf{Method} & \textbf{GAUC} & \textbf{Relative Lift} \\
\midrule
DIN              & 0.6006 (3E-5) & --     \\
TWIN             & 0.6079 (6E-5) & +1.22\% \\
MISS             & 0.6087 (1E-5) & +1.35\% \\
\midrule
SIM-Hard         & 0.6145 (6E-5) & +2.31\% \\
SIM-Soft         & 0.6144 (5E-5) & +2.30\% \\
MUSE             & 0.6148 (7E-5)  & +2.36\% \\
\midrule
\SemIDGSU        & 0.6148 (7E-5) & +2.36\% \\
\textbf{\SimGSU} & \textbf{0.6155 (9E-5)} & \textbf{+2.48\%} \\
\bottomrule
\end{tabular}
\end{table}

\textbf{Lifelong behavior modeling consistently outperforms short-behavior modeling.} DIN, which operates only on recent behaviors, yields the lowest GAUC of 0.6006. In contrast, all two-stage methods leveraging lifelong behavior sequences via GSU retrieval achieve substantial improvements, underscoring the necessity of modeling long-term user interests in industrial recommendation settings.

\textbf{Multi-modal methods outperform ID-only baselines.} ID-centric lifelong models (TWIN, MISS, SIM-Hard, and SIM-Soft) underperform compared to multi-modal approaches. Specifically, TWIN's ID-based retrieval generalizes poorly to long-tail items, while MISS and the SIM variants lack sufficient multi-modal integration in the ESU stage, placing them below \model.

\textbf{\model achieves the best overall performance.} 
\model achieves the highest GAUC of 0.6155, outperforming the strong baseline MUSE by 0.11\%. 
Although the absolute gain over MUSE is modest, AUC/GAUC improvements at the 0.1\% level are widely regarded as practically meaningful in large-scale recommendation and CTR prediction systems~\cite{ETA,fan2022racp,dare}.
Notably, \SemIDGSU achieves comparable performance to MUSE while replacing dense similarity-based retrieval with efficient SemID-based lookup, demonstrating that SemID can serve as an effective and scalable surrogate for multi-modal similarity.
The performance gain of the full model stems from \model's unified target-conditioned framework, where prefix-encoded SemID and similarity-aware bucket are directly integrated into sequential modeling.

\subsection{Ablation Study}
\label{sec:ablation}
To understand the contribution of each component, we conduct ablation experiments by progressively adding components to the base model. Results are reported in Table~\ref{tab:ablation}.

\begin{table}[t]
\centering
\caption{Ablation study on GAUC. All variants use similarity-based soft retrieval as the GSU strategy. The base model uses target attention with only ID-based representations over the retrieved lifelong behavior sequence. TI denotes Target-conditioned Interaction.}
\label{tab:ablation}
\begin{tabular}{lcc}
\toprule
\textbf{ESU Configuration} & \textbf{GAUC} & \textbf{Relative Lift} \\
\midrule
Target Attention (base)           & 0.6080 & --     \\
+ SemID only              & 0.6095 & +0.25\% \\
+ SimBucket only          & 0.6142 & +1.02\% \\
+ SimBucket + SemID& 0.6153 & +1.20\% \\
\textbf{\model (SimBucket + SemID + TI)} & \textbf{0.6155} & \textbf{+1.23\%} \\
\bottomrule
\end{tabular}
\end{table}

\begin{figure*}[!t]
\centering
\begin{minipage}[t]{0.70\textwidth}
    \centering
    \includegraphics[width=\linewidth]{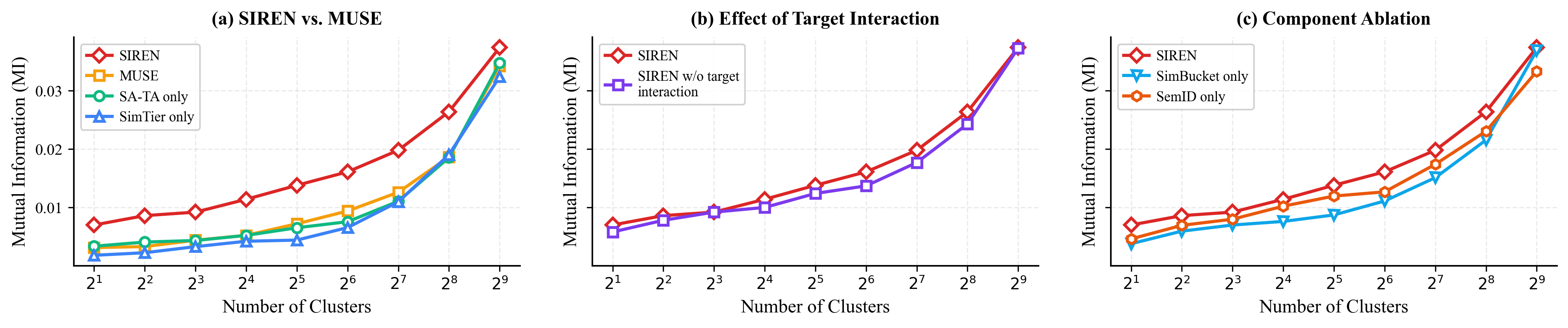}
\end{minipage}
\hfill
\begin{minipage}[t]{0.27\textwidth}
    \centering
    \includegraphics[width=\linewidth]{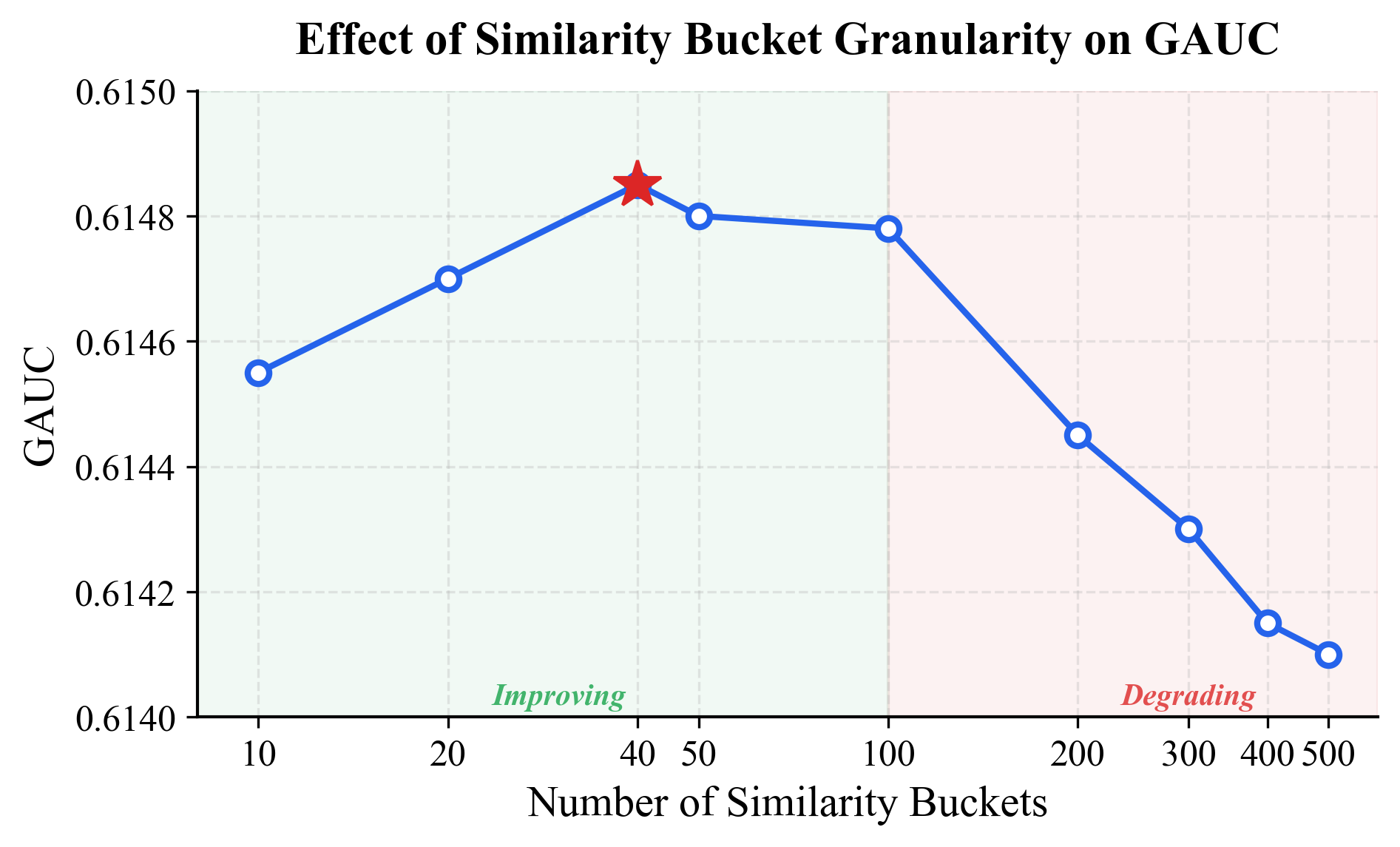}
\end{minipage}
\vspace{-1mm}
\caption{
Analysis of representation discriminability and similarity-bucket granularity.
Left: representation discriminability measured by mutual information between clustered user interest representations and click labels, including comparisons between \model and MUSE, the effect of target-aware interaction, and the ablation of SimBucket and SemID.
Right: effect of similarity-bucket granularity on GAUC, where performance peaks at 40 buckets and degrades with further increases in granularity.
}
\label{fig:analysis}
\vspace{-3mm}
\end{figure*}

\textbf{Both side-information contribute positively.}
Adding SemID alone improves GAUC from 0.6080 to 0.6095, while similarity buckets alone yield a larger gain to 0.6142.
Their combination further improves performance to 0.6153, exceeding either component in isolation and demonstrating their complementary effects. And the Target-conditioned Interaction provides additional gain of GAUC. Although the performance gain is modest, the mutual information analysis in Fig.~\ref{fig:analysis} shows that it substantially enhances representation discriminability.

\textbf{Similarity buckets provide stronger gains, while SemIDs refine semantic distinctions.}
The larger improvement from similarity buckets is expected, as they directly encode target-aware relevance, which is the most informative signal for interest modeling.
In contrast, SemIDs capture target-independent semantic content, helping distinguish behaviors with similar similarity scores but different semantics.
Together, they jointly model \emph{how relevant} a behavior is to the target and \emph{what it represents}, enabling more comprehensive interest modeling.

% \begin{figure*}[!ht]
% \centering
% \includegraphics[width=\textwidth]{images/fig_mi_comparison.png}
% \caption{Representation discriminability measured by mutual information between clustered user interest representations and click labels. (a)~\model vs.\ MUSE and its components; (b)~Effect of target-aware interaction; (c)~Component ablation of SimBucket and SemID.}
% \label{fig:mi}
% \end{figure*}

% \begin{figure}[t]
% \centering
% \includegraphics[width=0.75\columnwidth]{images/fig_bucket_gauc.png}
% \caption{Effect of similarity bucket granularity on GAUC. Performance peaks at 40 buckets and degrades with further increases.}
% \label{fig:bucket_gauc}
% \vspace{-10}
% \end{figure}

\subsection{Representation Discriminability}
\label{sec:discriminability}
Beyond performance metrics, we further investigate whether \model learns more informative and discriminative user representations. To this end, we evaluate their mutual information (MI) with the click label.

Specifically, we employ the user interest representation $\tilde{u}_t$ defined in Eq.~(\ref{eq:target_aware_user_interest}).
Since $\tilde{u}_t$ is continuous and high-dimensional, we first apply $K$-means clustering to quantize the representation space, assigning each $\tilde{u}_t$ to a discrete cluster $Q(\tilde{u}_t)$. 
We then compute the MI between the cluster assignment $Q(\tilde{u}_t)$ and the binary click label $Y$.
This metric serves as a proxy for representation discriminability: a higher MI indicates that the learned representations more effectively separate positive and negative samples.

As shown in Fig.~\ref{fig:analysis}(left), we compare representation discriminability across different model configurations and draw the following observations.

\textbf{\model consistently produces more discriminative representations.} 
Across varying numbers of $K$-means clusters, \model achieves consistently higher MI than the MUSE interest representation, as well as its two individual multi-modal components: \emph{SimTier}, which aggregates target-behavior similarity into a global sequence-level histogram, and \emph{SA-TA} (Semantic-Aware Target Attention), which infuses semantic similarities into ID-based attention weights.
This improvement can be attributed to \model's unified sequential modeling framework with coarse and fine-grained multi-modal signals.

\textbf{Target-conditioned interaction significantly enhances representation quality.} 
As shown in Fig.~\ref{fig:analysis}(b), enabling the target interaction mechanism clearly improves MI. This demonstrates that explicitly modeling the correlation between the extracted user interest and target representation enhances feature interaction and strengthens the discriminability of user representations.

\textbf{Similarity buckets and SemIDs provide complementary information.} 
Fig.~\ref{fig:analysis}(c) shows that combining similarity buckets and SemIDs achieves the highest MI, outperforming either one. 
This confirms that the two types of side information capture complementary signals of user behavior: similarity buckets encode target-aware relevance, while SemIDs capture item semantic information. 
Their joint integration enables the model to achieve a more effective fusion of multi-modal features.

Overall, these results provide consistent evidence that \model improves not only predictive performance but also the intrinsic quality of learned user representations.

\subsection{Necessity of Fine-Grained Semantic IDs}
\label{sec:semid_analysis}

We further analyze why fine-grained Semantic IDs are necessary beyond coarse similarity buckets. Prefix-encoded SemIDs and similarity buckets encode multi-modal information at different granularities: the former captures item-level semantic structure, while the latter provides target-conditioned relevance. We examine their discriminative power and complementarity from three perspectives.

\subsubsection{Conditional Entropy Analysis.}
We first evaluate how different grouping strategies reduce click-label uncertainty. Given a grouping variable $G$, we compute the information gain
\begin{equation}
    I(Y;G)=H(Y)-H(Y|G),
\end{equation}
where $Y$ denotes the click label. A larger $I(Y;G)$ indicates that the grouping better explains label variation and thus induces more discriminative user-interest partitions.

Semantic-ID grouping achieves $I(Y;\mathrm{SID})=0.0195$, which is substantially higher than the information gain of similarity-bucket grouping, $I(Y;\mathrm{Sim})=0.0056$. 
This result shows that SemIDs preserve more click-related information than scalar similarity buckets, suggesting that they provide finer-grained partitions that better align with user feedback.

\subsubsection{Within-Bucket CTR Distribution Analysis.}
We further examine whether similarity buckets can sufficiently distinguish behavior--target pairs with similar multi-modal proximity. 
As shown in Figure~\ref{fig:teaser_coarsefine}, even within the same similarity bucket, CTRs vary substantially across Semantic ID groups. 
Moreover, this within-bucket dispersion becomes larger in higher-similarity regions: the CTR range increases from $[0.815,0.949]$ in bucket $[0.2,0.3)$ to $[0.755,0.956]$ in bucket $[0.5,0.6)$, and further to $[0.735,1.000]$ in bucket $[0.9,1.0]$. The corresponding standard deviation also increases from about $0.035$ to $0.052$ and $0.068$, respectively.

These results indicate that \emph{high target--behavior similarity does not imply homogeneous user responses}. Behavior--target pairs that are close in the multi-modal space can still differ significantly in the collaborative CTR space. 
Therefore, \emph{coarse similarity buckets alone are insufficient to capture the fine-grained heterogeneity required for accurate interest modeling, especially in high-similarity regions}.

\subsubsection{Limitation of Increasing Bucket Granularity.}
A natural question is whether using more similarity buckets can recover the missing information. To answer this, we evaluate MUSE with different bucket granularities.

As shown in Fig.~\ref{fig:analysis}(right), increasing the number of buckets from 20 to 40 improves GAUC, but further increasing the granularity leads to performance degradation. This suggests that the limitation is not merely caused by insufficient bucket resolution. Instead, histogram-style similarity summarization compresses item-level similarity sequences into global statistics, discarding item identity and temporal structure. In contrast, \model preserves both similarity and SemID signals at the item level within the ESU.

Overall, these analyses show that similarity buckets and SemIDs are complementary: similarity buckets encode coarse target-conditioned relevance, while SemIDs capture fine-grained semantic and collaborative heterogeneity. Such complementarity cannot be recovered by simply refining similarity bucket granularity.

\subsection{Online A/B Tests}
\label{sec:online}
We deployed \model in Tencent's Weixin online advertising system, one of the largest online advertising platforms serving tens of billions of ad requests per day across multiple production scenarios. To capture lifelong user interests, the production model contains cross-domain behavior sequences spanning advertisements, Channels, and content feeds, covering up to two years of user interactions with a maximum sequence length of 4,000 per domain.

We conduct A/B tests across three major advertising scenarios with 20\% of traffic over a maximum period of 14 days per experiment. The production baseline was based on SIM-Hard~\cite{pi2020sim}. All reported improvements have been verified through rigorous significance testing under varying traffic ratios, and \model has since been fully deployed across all evaluated scenarios. Table~\ref{tab:online} summarizes the online results.

\model consistently improves GMV across all three scenarios, achieving lifts of +2.28\% on Moments, +3.87\% on Official Accounts, and +1.61\% on Channels. The gains span multiple pipeline stages, demonstrating that the proposed framework generalizes well to diverse real scenarios.

\begin{table}[t]
\centering
\caption{Online A/B test results of \model across different advertising scenarios and pipeline stages. }
\label{tab:online}
\begin{tabular}{llll}
\toprule
\textbf{Scenario} & \textbf{Stage} & \textbf{Feature} & \textbf{GMV Lift} \\
\midrule
\multirow{2}{*}{Weixin Moments}
  & pCTR   & Similarity & +1.58\% \\
  & pCTR   & SemID     & +0.70\% \\
\midrule
\multirow{2}{*}{Weixin Official Accounts}
  & pCTR   & Similarity & +1.64\% \\
  & pCVR   & SemID     & +2.23\% \\
\midrule
\multirow{2}{*}{Weixin Channels}
  & pCTR      & Similarity & +0.87\% \\
  & LTR & SemID     & +0.74\% \\
\bottomrule
\end{tabular}
\end{table}

\paragraph{Cold-start analysis}
To better understand where the gains originate, we perform a fine-grained analysis over user activity levels and ad cold-start scenarios on Weixin Moments. As shown in Table~\ref{tab:coldstart}, the GMV lift increases monotonically as user activity decreases, with \model delivering pronounced gains for both \emph{low-activity users} (<50 interactions) and \emph{cold-start users} (<10 interactions). In these scenarios where sparse interaction data renders traditional ID-based signals unreliable, our multi-modal side information substantially improves interest modeling. Similarly, on the item side, \model effectively leverages visual and textual content to enhance interest matching, thereby alleviating the cold-start problem for new ads launched within the past 24 hours.

\begin{table}[t]
\centering
\caption{Relative GMV lift compared to the overall average on Weixin Moments.}
\label{tab:coldstart}
\begin{tabular}{llc}
\toprule
\textbf{Side} & \textbf{Segment} & \textbf{Relative Lift vs.\ Overall} \\
\midrule
\multirow{2}{*}{User-side}
  & Low-activity users   & $\sim$1.7$\times$ \\
  & Cold-start users     & $\sim$3.6$\times$ \\
\midrule
Item-side
  & New ads (first day)  & $\sim$1.4$\times$ \\
\bottomrule
\end{tabular}
\end{table}

\paragraph{Deployment efficiency of SemID-based hard retrieval.}
We further analyze SemID-based as an alternative retrieval strategy in the GSU stage.
Compared to similarity-based retrieval, which requires maintaining dense embedding indices and computing pairwise similarities over high-dimensional vectors, SemID-based retrieval significantly reduces online serving cost by over 90\% in both latency and storage.
Importantly, this efficiency gain comes with minimal performance degradation.
These results suggest that SemID-based strategy provides a practical efficiency–performance trade-off, offering a cost-effective alternative for large-scale deployment.

\section{Conclusion}
In this paper, we propose \model, a unified multi-granularity semantic interaction framework for multi-modal lifelong user interest modeling. In the GSU stage, similarity-based soft retrieval and SemID-based hard retrieval are explored to provide a flexible trade-off between retrieval effectiveness and serving efficiency. In the ESU stage, prefix-encoded SemIDs and similarity buckets are incorporated as complementary multi-granular side information, enabling unified target-conditioned sequential modeling alongside ID-based features.
Experimental results on both offline benchmarks and large-scale real-world online systems demonstrate that \model effectively integrates  multi-modal signals into lifelong interest modeling, and significantly enhance recommendation performance in industrial scenarios.

\newpage
\bibliographystyle{ACM-Reference-Format}
\bibliography{refs}

@inproceedings{fan2022racp,
  title={Modeling Users' Contextualized Page-wise Feedback for Click-Through Rate Prediction in E-commerce Search},
  author={Fan, Zhifang and Ou, Dan and Gu, Yulong and Fu, Bairan and Li, Xiang and Bao, Wentian and Dai, Xin-Yu and Zeng, Xiaoyi and Zhuang, Tao and Liu, Qingwen},
  booktitle={Proceedings of the Fifteenth ACM International Conference on Web Search and Data Mining},
  pages={262--270},
  year={2022}
}

@inproceedings{zhou2018din,
  author    = {Guorui Zhou and Xiaoqiang Zhu and Chenru Song and Ying Fan and Han Zhu and Xiao Ma and Yanghui Yan and Junqi Jin and Han Li and Kun Gai},
  title     = {Deep Interest Network for Click-Through Rate Prediction},
  booktitle = {KDD},
  year      = {2018}
}

@inproceedings{UBM,
  author       = {Zhicheng He and
                  Weiwen Liu and
                  Wei Guo and
                  Jiarui Qin and
                  Yingxue Zhang and
                  Yaochen Hu and
                  Ruiming Tang},
  title        = {A Survey on User Behavior Modeling in Recommender Systems},
  booktitle    = {{IJCAI}},
  pages        = {6656--6664},
  publisher    = {ijcai.org},
  year         = {2023}
}

@article{ultra-hstu,
  author       = {Qin Ding and
                  Kevin Course and
                  Linjian Ma and
                  Jianhui Sun and
                  Ruochen Liu and
                  Zhao Zhu and
                  Chunxing Yin and
                  Wei Li and
                  Dai Li and
                  Yu Shi and
                  Xuan Cao and
                  Ze Yang and
                  Han Li and
                  Xing Liu and
                  Bi Xue and
                  Hongwei Li and
                  Rui Jian and
                  Daisy Shi He and
                  Jing Qian and
                  Matt Ma and
                  Qunshu Zhang and
                  Rui Li},
  title        = {Bending the Scaling Law Curve in Large-Scale Recommendation Systems},
  journal      = {arXiv Preprint},
  volume       = {https://arxiv.org/abs/2602.16986},
  year         = {2026}
}

@article{pi2020sim,
  author  = {Qi Pi and Guorui Zhou and Yujing Zhang and Zhe Wang and Lejian Ren and Ying Fan and Xiaoqiang Zhu and Kun Gai},
  title   = {Search-based User Interest Modeling with Lifelong Sequential Behavior Data for Click-Through Rate Prediction},
  journal = {arXiv Preprint},
  url = {https://arxiv.org/abs/2006.05639},
  year    = {2020}
}

@article{ETA,
  author       = {Qiwei Chen and
                  Changhua Pei and
                  Shanshan Lv and
                  Chao Li and
                  Junfeng Ge and
                  Wenwu Ou},
  title        = {End-to-End User Behavior Retrieval in Click-Through RatePrediction
                  Model},
  journal      = {arXiv Preprint},
  volume       = {https://arxiv.org/abs/2108.04468},
  year         = {2021}
}

@inproceedings{twin,
  author       = {Jianxin Chang and
                  Chenbin Zhang and
                  Zhiyi Fu and
                  Xiaoxue Zang and
                  Lin Guan and
                  Jing Lu and
                  Yiqun Hui and
                  Dewei Leng and
                  Yanan Niu and
                  Yang Song and
                  Kun Gai},
  title        = {{TWIN:} TWo-stage Interest Network for Lifelong User Behavior Modeling
                  in {CTR} Prediction at Kuaishou},
  booktitle    = {{KDD}},
  pages        = {3785--3794},
  year         = {2023}
}

@inproceedings{twinv2,
  author       = {Zihua Si and
                  Lin Guan and
                  Zhongxiang Sun and
                  Xiaoxue Zang and
                  Jing Lu and
                  Yiqun Hui and
                  Xingchao Cao and
                  Zeyu Yang and
                  Yichen Zheng and
                  Dewei Leng and
                  Kai Zheng and
                  Chenbin Zhang and
                  Yanan Niu and
                  Yang Song and
                  Kun Gai},
  title        = {{TWIN} {V2:} Scaling Ultra-Long User Behavior Sequence Modeling for
                  Enhanced {CTR} Prediction at Kuaishou},
  booktitle    = {{CIKM}},
  pages        = {4890--4897},
  year         = {2024}
}

@inproceedings{longer,
  author       = {Zheng Chai and
                  Qin Ren and
                  Xijun Xiao and
                  Huizhi Yang and
                  Bo Han and
                  Sijun Zhang and
                  Di Chen and
                  Hui Lu and
                  Wenlin Zhao and
                  Lele Yu and
                  Xionghang Xie and
                  Shiru Ren and
                  Xiang Sun and
                  Yaocheng Tan and
                  Peng Xu and
                  Yuchao Zheng and
                  Di Wu},
  title        = {{LONGER:} Scaling Up Long Sequence Modeling in Industrial Recommenders},
  booktitle    = {RecSys},
  pages        = {247--256},
  year         = {2025}
}

@article{stca,
  author       = {Lin Guan and
                  Jia{-}Qi Yang and
                  Zhishan Zhao and
                  Beichuan Zhang and
                  Bo Sun and
                  Xuanyuan Luo and
                  Jinan Ni and
                  Xiaowen Li and
                  Yuhang Qi and
                  Zhifang Fan and
                  Hangyu Wang and
                  Qiwei Chen and
                  Yi Cheng and
                  Feng Zhang and
                  Xiao Yang},
  title        = {Make It Long, Keep It Fast: End-to-End 10k-Sequence Modeling at Billion
                  Scale on Douyin},
  journal      = {arXiv Preprint},
  volume       = {https://arxiv.org/abs/2511.06077},
  year         = {2025}
}

@inproceedings{dare,
  author       = {Ningya Feng and
                  Junwei Pan and
                  Jialong Wu and
                  Baixu Chen and
                  Ximei Wang and
                  Qian Li and
                  Xian Hu and
                  Jie Jiang and
                  Mingsheng Long},
  title        = {Long-Sequence Recommendation Models Need Decoupled Embeddings},
  booktitle    = {{ICLR}},
  year         = {2025}
}

@article{tencent-long,
  author       = {Xian Hu and
                  Ming Yue and
                  Zhixiang Feng and
                  Junwei Pan and
                  Junjie Zhai and
                  Ximei Wang and
                  Xinrui Miao and
                  Qian Li and
                  Xun Liu and
                  Shangyu Zhang and
                  Letian Wang and
                  Hua Lu and
                  Zijian Zeng and
                  Chen Cai and
                  Wei Wang and
                  Fei Xiong and
                  Pengfei Xiong and
                  Jintao Zhang and
                  Zhiyuan Wu and
                  Chunhui Zhang and
                  Anan Liu and
                  Jiulong You and
                  Chao Deng and
                  Yuekui Yang and
                  Shudong Huang and
                  Dapeng Liu and
                  Haijie Gu},
  title        = {Practice on Long Behavior Sequence Modeling in Tencent Advertising},
  journal      = {arXiv Preprint},
  volume       = {https://arxiv.org/abs/2510.21714},
  year         = {2025}
}

@inproceedings{lcn,
  author       = {Ruijie Hou and
                  Zhaoyang Yang and
                  Ming Yu and
                  Hongyu Lu and
                  Zhuobin Zheng and
                  Yu Chen and
                  Qinsong Zeng and
                  Ming Chen},
  title        = {Cross-Domain LifeLong Sequential Modeling for Online Click-Through
                  Rate Prediction},
  booktitle    = {{KDD}},
  pages        = {5116--5125},
  year         = {2024}
}

@inproceedings{tin,
  author       = {Haolin Zhou and
                  Junwei Pan and
                  Xinyi Zhou and
                  Xihua Chen and
                  Jie Jiang and
                  Xiaofeng Gao and
                  Guihai Chen},
  title        = {Temporal Interest Network for User Response Prediction},
  booktitle    = {{WWW}},
  pages        = {413--422},
  year         = {2024}
}

@article{MMRecSurvey2,
  author       = {Alejo Lopez{-}Avila and
                  Jinhua Du},
  title        = {A Survey on Large Language Models in Multimodal Recommender Systems},
  journal      = {arXiv Preprint},
  volume       = {https://arxiv.org/abs/2505.09777},
  year         = {2025}
}

@article{MMRecSurvey1,
  author       = {Qidong Liu and
                  Jiaxi Hu and
                  Yutian Xiao and
                  Xiangyu Zhao and
                  Jingtong Gao and
                  Wanyu Wang and
                  Qing Li and
                  Jiliang Tang},
  title        = {Multimodal Recommender Systems: {A} Survey},
  journal      = {{ACM} Comput. Surv.},
  volume       = {57},
  number       = {2},
  pages        = {26:1--26:17},
  year         = {2025}
}

@inproceedings{Clip,
  author       = {Alec Radford and
                  Jong Wook Kim and
                  Chris Hallacy and
                  Aditya Ramesh and
                  Gabriel Goh and
                  Sandhini Agarwal and
                  Girish Sastry and
                  Amanda Askell and
                  Pamela Mishkin and
                  Jack Clark and
                  Gretchen Krueger and
                  Ilya Sutskever},
  title        = {Learning Transferable Visual Models From Natural Language Supervision},
  booktitle    = {{ICML}},
  pages        = {8748--8763},
  year         = {2021}
}

@inproceedings{MMLLMSurvey,
  author       = {Davide Caffagni and
                  Federico Cocchi and
                  Luca Barsellotti and
                  Nicholas Moratelli and
                  Sara Sarto and
                  Lorenzo Baraldi and
                  Marcella Cornia and
                  Rita Cucchiara},
  title        = {The Revolution of Multimodal Large Language Models: {A} Survey},
  booktitle    = {{ACL}},
  pages        = {13590--13618},
  year         = {2024}
}

@article{Qwen3-vl,
  author       = {Qwen Team},
  title        = {Qwen3-VL Technical Report},
  journal      = {arXiv Preprint},
  volume       = {https://arxiv.org/abs/2511.21631},
  year         = {2025}
}

@inproceedings{MISSRec,
  author       = {Jinpeng Wang and
                  Ziyun Zeng and
                  Yunxiao Wang and
                  Yuting Wang and
                  Xingyu Lu and
                  Tianxiang Li and
                  Jun Yuan and
                  Rui Zhang and
                  Hai{-}Tao Zheng and
                  Shu{-}Tao Xia},
  title        = {MISSRec: Pre-training and Transferring Multi-modal Interest-aware
                  Sequence Representation for Recommendation},
  booktitle    = {{ACM} Multimedia},
  year         = {2023}
}

@inproceedings{HSTU,
  author       = {Jiaqi Zhai and
                  Lucy Liao and
                  Xing Liu and
                  Yueming Wang and
                  Rui Li and
                  Xuan Cao and
                  Leon Gao and
                  Zhaojie Gong and
                  Fangda Gu and
                  Jiayuan He and
                  Yinghai Lu and
                  Yu Shi},
  title        = {Actions Speak Louder than Words: Trillion-Parameter Sequential Transducers
                  for Generative Recommendations},
  booktitle    = {{ICML}},
  pages        = {58484--58509},
  year         = {2024}
}

@inproceedings{AlignRec,
  author       = {Yifan Liu and
                  Kangning Zhang and
                  Xiangyuan Ren and
                  Yanhua Huang and
                  Jiarui Jin and
                  Yingjie Qin and
                  Ruilong Su and
                  Ruiwen Xu and
                  Yong Yu and
                  Weinan Zhang},
  title        = {AlignRec: Aligning and Training in Multimodal Recommendations},
  booktitle    = {{CIKM}},
  pages        = {1503--1512},
  year         = {2024}
}

@inproceedings{SimTier,
  author       = {Xiang{-}Rong Sheng and
                  Feifan Yang and
                  Litong Gong and
                  Biao Wang and
                  Zhangming Chan and
                  Yujing Zhang and
                  Yueyao Cheng and
                  Yong{-}Nan Zhu and
                  Tiezheng Ge and
                  Han Zhu and
                  Yuning Jiang and
                  Jian Xu and
                  Bo Zheng},
  title        = {Enhancing Taobao Display Advertising with Multimodal Representations:
                  Challenges, Approaches and Insights},
  booktitle    = {{CIKM}},
  pages        = {4858--4865},
  year         = {2024}
}

@inproceedings{QARM,
  author       = {Xinchen Luo and
                  Jiangxia Cao and
                  Tianyu Sun and
                  Jinkai Yu and
                  Rui Huang and
                  Wei Yuan and
                  Hezheng Lin and
                  Yichen Zheng and
                  Shiyao Wang and
                  Qigen Hu and
                  Changqing Qiu and
                  Jiaqi Zhang and
                  Xu Zhang and
                  Zhiheng Yan and
                  Jingming Zhang and
                  Simin Zhang and
                  Mingxing Wen and
                  Zhaojie Liu and
                  Guorui Zhou},
  title        = {{QARM:} Quantitative Alignment Multi-Modal Recommendation at Kuaishou},
  booktitle    = {{CIKM}},
  pages        = {5915--5922},
  year         = {2025}
}

@inproceedings{MISS,
  author       = {Chengcheng Guo and
                  Junda She and
                  Kuo Cai and
                  Shiyao Wang and
                  Qigen Hu and
                  Qiang Luo and
                  Guorui Zhou and
                  Kun Gai},
  title        = {{MISS:} Multi-Modal Tree Indexing and Searching with Lifelong Sequential
                  Behavior for Retrieval Recommendation},
  booktitle    = {{CIKM}},
  pages        = {5683--5690},
  year         = {2025}
}

@article{muse,
  author       = {Bin Wu and
                  Feifan Yang and
                  Zhangming Chan and
                  Yu{-}Ran Gu and
                  Jiawei Feng and
                  Chao Yi and
                  Xiang{-}Rong Sheng and
                  Han Zhu and
                  Jian Xu and
                  Mang Ye and
                  Bo Zheng},
  title        = {{MUSE:} {A} Simple Yet Effective Multimodal Search-Based Framework
                  for Lifelong User Interest Modeling},
  journal      = {arXiv Preprint},
  volume       = {https://arxiv.org/abs/2512.07216},
  year         = {2025}
}

@inproceedings{letter,
  author       = {Wenjie Wang and
                  Honghui Bao and
                  Xinyu Lin and
                  Jizhi Zhang and
                  Yongqi Li and
                  Fuli Feng and
                  See{-}Kiong Ng and
                  Tat{-}Seng Chua},
  title        = {Learnable Item Tokenization for Generative Recommendation},
  booktitle    = {{CIKM}},
  pages        = {2400--2409},
  year         = {2024}
}

@inproceedings{DAS,
  author       = {Wencai Ye and
                  Mingjie Sun and
                  Shaoyun Shi and
                  Peng Wang and
                  Wenjin Wu and
                  Peng Jiang},
  title        = {{DAS:} Dual-Aligned Semantic IDs Empowered Industrial Recommender
                  System},
  booktitle    = {{CIKM}},
  pages        = {6217--6224},
  year         = {2025}
}

@inproceedings{MMQ,
  author       = {Yi Xu and
                  Moyu Zhang and
                  Chenxuan Li and
                  Zhihao Liao and
                  Haibo Xing and
                  Hao Deng and
                  Jinxin Hu and
                  Yu Zhang and
                  Xiaoyi Zeng and
                  Jing Zhang},
  title        = {{MMQ:} Multimodal Mixture-of-Quantization Tokenization for Semantic
                  {ID} Generation and User Behavioral Adaptation},
  booktitle    = {{WSDM}},
  pages        = {788--797},
  year         = {2026}
}

@inproceedings{prefix-sid,
  author       = {Carolina Zheng and
                  Minhui Huang and
                  Dmitrii Pedchenko and
                  Kaushik Rangadurai and
                  Siyu Wang and
                  Fan Xia and
                  Gaby Nahum and
                  Jie Lei and
                  Yang Yang and
                  Tao Liu and
                  Zutian Luo and
                  Xiaohan Wei and
                  Dinesh Ramasamy and
                  Jiyan Yang and
                  Yiping Han and
                  Lin Yang and
                  Hangjun Xu and
                  Rong Jin and
                  Shuang Yang},
  title        = {Enhancing Embedding Representation Stability in Recommendation Systems
                  with Semantic {ID}},
  booktitle    = {RecSys},
  pages        = {954--957},
  year         = {2025}
}

@article{RQ-VAE,
  author       = {Neil Zeghidour and
                  Alejandro Luebs and
                  Ahmed Omran and
                  Jan Skoglund and
                  Marco Tagliasacchi},
  title        = {SoundStream: An End-to-End Neural Audio Codec},
  journal      = {{IEEE} {ACM} Trans. Audio Speech Lang. Process.},
  volume       = {30},
  pages        = {495--507}
  ,
  year         = {2022}
}

@inproceedings{AdamW,
  author       = {Ilya Loshchilov and
                  Frank Hutter},
  title        = {Decoupled Weight Decay Regularization},
  booktitle    = {{ICLR}},
  year         = {2019}
}
\end{document}